\documentclass[sigconf,balance=False, pbalance]{acmart}
\usepackage{tcolorbox}
\usepackage{xspace}
\usepackage{enumitem}
\usepackage{caption}
\usepackage{subcaption}
\usepackage{bbding}
\usepackage{url}
\usepackage{booktabs}
\usepackage{tabularx}
\usepackage{circledsteps} 
\usepackage{graphicx}
\usepackage{footmisc}

\usepackage{caption}

\usepackage{multicol}

\usepackage{makecell}
\usepackage{multirow}
\newcolumntype{C}[1]{>{\centering\arraybackslash}p{#1}}

\usepackage{amsmath,bm,amsthm}
\theoremstyle{definition}

\usepackage{algorithm}
\usepackage{algpseudocode}
\algnewcommand\algorithmicforeach{\textbf{for each}}
\algdef{SE}[FUNCTION]{Function}{EndFunction}[2]{\algorithmicfunction\ \textnormal{#1}\ (#2)}{\algorithmicend\ \algorithmicfunction}%
\algdef{S}[FOR]{ForEach}[1]{\algorithmicforeach\ #1\ \algorithmicdo}
\algrenewcommand\alglinenumber[1]{\footnotesize #1}
\algrenewcommand\algorithmicrequire{\small \textbf{input:}}
\algrenewcommand\algorithmicensure{\small \textbf{output:}}
\algrenewcommand\algorithmicfunction{\textbf{Function}}
\algtext*{EndFunction} %
\algtext*{EndIf} %
\algtext*{EndFor} %
\algtext*{EndWhile} %

\usepackage{fontawesome} 
\usepackage{framed}
\usepackage{textcomp}

\usepackage{microtype}
\usepackage{ifthen}

\newcounter{rq}
\newenvironment{answertorq}{%
    \begin{tcolorbox}[
            arc=2mm,
            boxrule=0.5pt,
            left=2pt,
            right=2pt,
            top=2pt,
            bottom=2pt
        ]
    \textbf{Answer to RQ\refstepcounter{rq}\therq{}:}
}{\end{tcolorbox}}

\tcbuselibrary{breakable}

\ifthenelse{\isundefined{\editmode}}{
    \newcommand{\editnote}[3]{%
    }
}
{
    \newcommand{\editnote}[3]{\xspace\colorbox{#1}{\sffamily \smaller \textcolor{white}{~\faCommenting{}~#2~}}\textcolor{#1}{~#3}\xspace}
}
\definecolor{latte-teal}{RGB}{23, 146, 153}
\definecolor{nord0}{HTML}{2E3440}
\definecolor{nord1}{HTML}{3B4252}
\definecolor{nord2}{HTML}{434C5E}
\definecolor{nord3}{HTML}{4C566A}
\definecolor{nord4}{HTML}{D8DEE9}
\definecolor{nord5}{HTML}{E5E9F0}
\definecolor{nord6}{HTML}{ECEFF4}
\definecolor{nord7}{HTML}{8FBCBB}
\definecolor{nord8}{HTML}{88C0D0}
\definecolor{nord9}{HTML}{81A1C1}
\definecolor{nord10}{HTML}{5E81AC}
\definecolor{nord11}{HTML}{BF616A}
\definecolor{nord12}{HTML}{D08770}
\definecolor{nord13}{HTML}{EBCB8B}
\definecolor{nord14}{HTML}{A3BE8C}
\definecolor{nord15}{HTML}{B48EAD}

\newcommand{\congying}[1]{\editnote{olive}{Congying}{#1}}

\newcommand{\scc}[1]{\editnote{violet}{S.C.}{#1}}
\newcommand{\jr}[1]{\editnote{brown}{jr}{#1}}
\newcommand{\hc}[1]{\editnote{latte-teal}{HC}{#1}}

\newcommand{\bella}[1]{\editnote{purple}{bella}{#1}}

\newcommand{\todo}[1]{\textcolor{black}{#1}}
\newcommand{\code}[1]{\texttt{#1}}
\newcommand{\tool}{\textsc{MR-Adopt}\xspace}

\newcommand{\editedA}[1]{\textcolor{black}{#1}}

\begin{document}
\copyrightyear{2024}
\acmYear{2024}
\setcopyright{acmlicensed}
\acmConference[ASE'24]{39th IEEE/ACM International Conference on Automated Software Engineering}{Oct 27 -- Nov 1, 2024}{Sacramento, California, United States}

\setcopyright{none}
\settopmatter{printacmref=false} 
\renewcommand\footnotetextcopyrightpermission[1]{} 



\title{\tool: Automatic Deduction of Input Transformation Function for Metamorphic Testing}


\author{{(This paper is accepted to ASE 2024)} \\ { ~ } }

\author{Congying Xu\textsuperscript{1,2}, Songqiang Chen\textsuperscript{1,2}, Jiarong Wu\textsuperscript{1,2}, Shing-Chi Cheung\textsuperscript{1,2*}, Valerio Terragni\textsuperscript{3}, Hengcheng Zhu\textsuperscript{1,2}, Jialun Cao\textsuperscript{1,2*}}
\affiliation{%
    \institution{\textsuperscript{1}The Hong Kong University of Science and Technology, Hong Kong, China}
    \country{}
}
\affiliation{%
    \institution{\textsuperscript{2}Guangzhou HKUST Fok Ying Tung Research Institute, China}
    \country{}
}
\affiliation{%
    \institution{\textsuperscript{3}The University of Auckland, Auckland, New Zealand}
    \country{}
}

\authornote{Shing-Chi Cheung and Jialun Cao are corresponding authors. Emails: scc@cse.ust.hk, jcaoap@connect.ust.hk}

\renewcommand{\shortauthors}{Congying Xu et al.}




\begin{abstract}
    While a recent study reveals that many developer-written test cases can encode a \todo{reusable} Metamorphic Relation (MR), over 70\% of them directly hard-code the source input and follow-up input in the encoded relation. 
Such encoded MRs, which do not contain an explicit input transformation to transform the source inputs to corresponding follow-up inputs, cannot be reused with new source inputs \todo{to enhance test adequacy}. 

In this paper, we propose \tool (\textit{\underline{A}utomatic \underline{D}eduction \underline{O}f in\underline{P}ut \underline{T}ransformation}) to automatically deduce the input transformation from the hard-coded source and follow-up inputs, aiming to enable the encoded MRs to be reused with new source inputs. 
With typically only one pair of source and follow-up inputs available in an MR-encoded test case as the example, we leveraged LLMs to understand the intention of the test case and generate additional examples of source-followup input pairs. 
This helps to guide the generation of input transformations generalizable to multiple source inputs.
Besides, to mitigate the issue that LLMs generate erroneous code, we refine LLM-generated transformations by removing MR-irrelevant code elements with data-flow analysis.
Finally, we assess candidate transformations based on encoded output relations and select the best transformation as the result.
Evaluation results show that \tool can generate input transformations 
applicable to all experimental source inputs for 72.00\% of encoded MRs, which is 33.33\% more than using vanilla GPT-3.5.
By incorporating \tool-generated input transformations, encoded MR-based test cases can effectively enhance the test adequacy, increasing the line coverage and mutation score by 10.62\% and 18.91\%, respectively. 

\end{abstract}

\maketitle


\newcommand{\wzd}{Wizard\xspace}
\newcommand{\gpt}{GPT-3.5\xspace}
\newcommand{\qw}{Qwen\xspace}
\newcommand{\lm}{Llama3\xspace}
\newcommand{\dps}{Deepseek\xspace}

\newcommand{\gptlong}{GPT-3.5-turbo-0125\xspace}
\newcommand{\wzdlong}{WizardCoder-15B\xspace}
\newcommand{\qwlong}{CodeQwen-1.5-7B\xspace}
\newcommand{\lmlong}{Llama3-8B-Instruct\xspace}
\newcommand{\dpslong}{Deepseek-coder-7b-instruct-v1.5\xspace}

\newcommand{\phaseOneIndex}{\textit{Phase1}\xspace}
\newcommand{\phaseTwoIndex}{\textit{Phase2}\xspace}
\newcommand{\phaseOne}{Input Pairs Preparation}
\newcommand{\phaseTwo}{Transformation Generation}

\newcommand{\phaseOneLong}{\textit{Phase 1:} Input Pair Preparation}
\newcommand{\phaseTwoLong}{\textit{Phase 2:} Transformation Generation}

\newcommand{\developerWrittenTestSuite}{$\mathcal{D}$}
\newcommand{\llmInputPairTestSuite}{$\mathcal{L}$}
\newcommand{\tranMRBasedTestSuite}{$\mathcal{M}$}

\newcommand{\relation}{\mathcal{R}}
\newcommand{\outputRelation}{\relation_o}
\newcommand{\inputRelation}{\relation_i}
\newcommand{\outputRelationWithElements}{\relation_o<\sourceInput, \followUpInput, \sourceOutput, \followUpOutput>}
\newcommand{\sourceInput}{x_{s}}
\newcommand{\followUpInput}{x_{f}}
\newcommand{\sourceOutput}{y_{s}}
\newcommand{\followUpOutput}{y_{f}}

\newcommand{\shortedSourceInputs}{\underset{v=1\cdots k}{\langle x_v \rangle}}
\newcommand{\shortedFollowUpInputs}{\underset{w=(k+1)\cdots n}{\langle x_w \rangle}}
\newcommand{\shortedAllInputs}{\underset{i=1\cdots n}{\langle x_i \rangle}}
\newcommand{\shortedSourceOutputs}{\underset{v=1\cdots k}{\langle f(x_v) \rangle}}
\newcommand{\shortedAllOutputs}{\underset{i=1\cdots n}{\langle f(x_i) \rangle}}

\newcommand{\PshortedSourceOutputs}{\underset{v=1\cdots k}{\langle P(x_v) \rangle}}
\newcommand{\PshortedAllOutputs}{\underset{i=1\cdots n}{\langle P(x_i) \rangle}}

\section{Introduction}

Metamorphic Testing (MT) is a powerful testing technique to address both the test case generation and the oracle problem~\cite{2016-segura-tse, chen2018metamorphic}.
Instead of assessing the outputs of individual inputs, MT validates the behavior of a subject under test (SUT) against Metamorphic Relations (MRs) for the SUT.
Each MR defines an \textit{input relation} over a set of related inputs and an \textit{output relation} over the expected outputs for those inputs. 
One \textit{outstanding benefit} of MT is that once an MR is identified, MT can leverage a wide range of automatically generated inputs (known as \emph{source inputs}) to exercise diverse program behaviors with no need to prepare oracles for individual inputs~\cite{mrscout}.
MT has achieved success in detecting faults for various software, such as {compilers \cite{MT4Compiler_PLDI14SZD,MT4Compiler_OOPSLA16SZD} and databases \cite{MT4DB_ICSE15,MT4DB_FSE24HPJ}.} 

Identifying appropriate MRs for a SUT is essential to applying MT.
Some studies have focused on MR identification.
Earlier approaches either suffer from being labor-intensive and specific to certain domains or pre-defined MR patterns~\cite{zhang2014search, zhang2019automatic, sun2016mumt} or produce overly generic MRs that are ineffective for testing, \todo{as well as recent LLM-based techniques}~\cite{DBLP:journals/corr/abs-2401-17019, DBLP:conf/iccS/TsigkanosRMK23}.
Recently, Xu et al.~\cite{mrscout} report that developers often encode domain knowledge in test cases that exercise MRs.
These encoded MRs can be generalized to many new inputs and serve as oracles for more exhaustive testing of the original programs (or programs with similar functionalities), by integrating with automatic input generation techniques~\cite{mrscout,2016-segura-tse, chen2018metamorphic}. 

\begin{figure}
\centering

\begin{subfigure}{0.46\textwidth}
\includegraphics[scale=0.41]{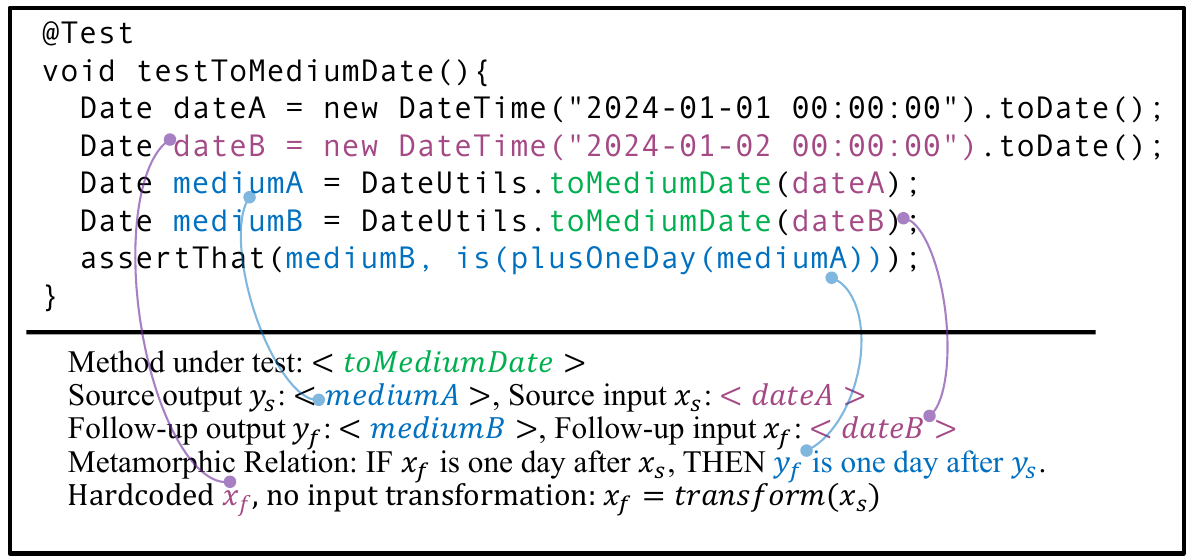}
\vspace{-16pt}
\caption{\footnotesize An MR-encoded test case (MTC) featuring a hardcoded follow-up input\footnotemark
 }
\label{fig:IT4MTa}
\end{subfigure}

\begin{subfigure}{0.46\textwidth}
\includegraphics[scale=0.41]{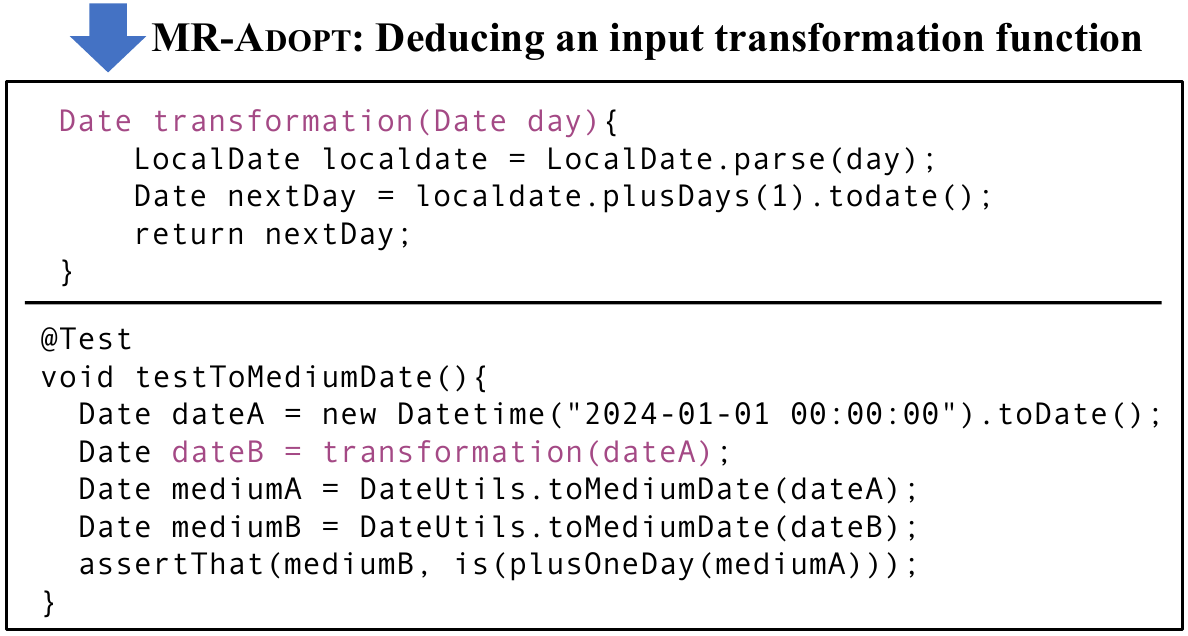}
\vspace{-16pt}
\caption{\footnotesize An MTC featuring a transformation-generated follow-up input}\label{fig:IT4MTb}
\vspace{3pt}
\end{subfigure}

\begin{subfigure}{0.46\textwidth}
\includegraphics[scale=0.41]{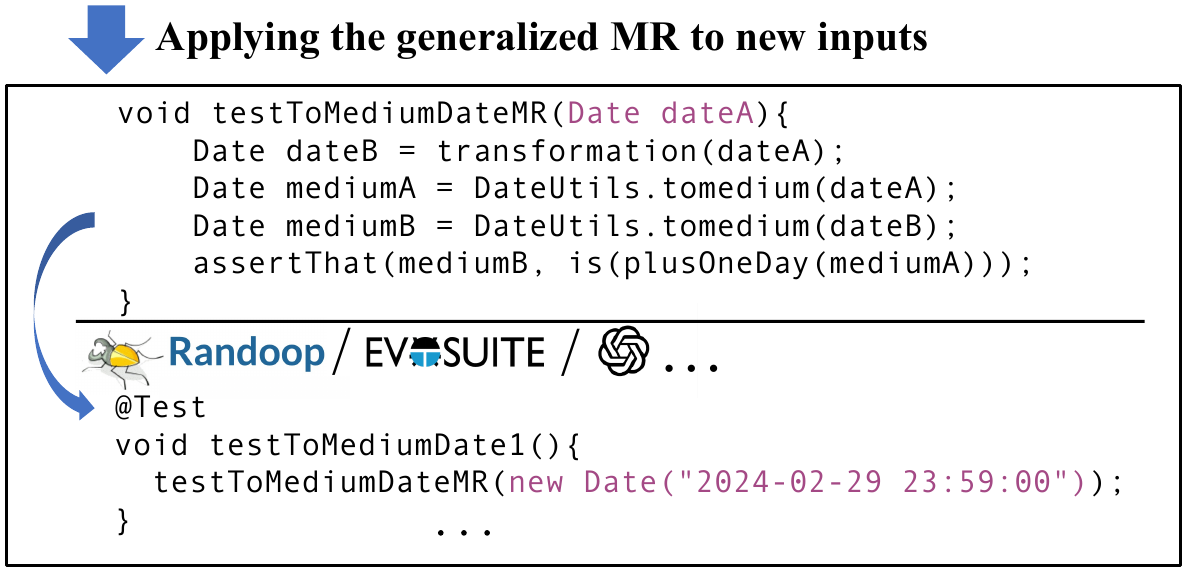}
\vspace{-16pt}
\caption{\footnotesize Metamorphic testing by integrating an MR with diverse source inputs}\label{fig:IT4MTc}
\end{subfigure}

\vspace{-5pt}
\caption{Overview of \small{\tool} \normalsize for Metamorphic Testing}
\label{fig:overview_of_IT4MT}
\end{figure}

    However, \citet{mrscout} show that over 70\% of \todo{11,000} MR-encoded test cases (MTCs) in their dataset do not contain explicit \todo{input relations}. 
    Instead, developers often hard-code the source and follow-up inputs.
    Figure~\ref{fig:IT4MTa} shows an MR-encoded test case intended to have the follow-up input (\code{dateB}) one day after the source input (\code{dateA}), but it simply hard-codes the two inputs.
    Without an explicit input transformation \todo{program}, follow-up inputs cannot be directly generated from automatically generated source inputs. This limitation hinders the reuse of valuable encoded MRs to achieve automated MT and enhance test adequacy. 
    \textbf{This paper aims to overcome this obstacle by inferring an explicit input relation from a given test case with its hard-coded input pairs.}
    Specifically, our goal is to construct an input transformation function that turns a source input into a follow-up input as shown in Figure~\ref{fig:IT4MTb}.
    With such input transformations, these encoded MRs can apply to a wider range of test inputs to test SUTs more exhaustively (Figure~\ref{fig:IT4MTc}).

This task can be viewed as a programming by example (PBE) problem, where the aim is to synthesize a transformation function that turns a given source input into the corresponding follow-up input. 
The challenge lies in \textit{correctly interpreting the contextual information, such as the relationship between hard-coded input pairs, output relations, and the properties of the SUT.} Moreover, with only one pair of source and follow-up inputs available as an example~\cite{mrscout}, there is a risk of generating program overfitted to the given example instead of realizing the true intention, as noted in existing PBE studies~\cite{DBLP:conf/popl/Gulwani11, DBLP:conf/icse/PanLNGLK21, DBLP:conf/fmcad/AlurBJMRSSSTU13}. 
Therefore, \textit{effectively leveraging contextual information is crucial to guide PBE and generate a generalizable input transformation} that aligns with the \textit{semantic} of the encoded MR, ensuring it applies to all potential source inputs with the corresponding output relation.

In this paper, we propose \tool, an approach that leverages large language models (LLMs) to automatically generate input transformation functions for MRs encoded in existing test cases.
Trained on extensive code corpora from various domains, LLMs have demonstrated effectiveness in code understanding~\cite{DBLP:conf/icse/NamMHVM24, DBLP:conf/icse/GengWD00JML24, DBLP:conf/icse/MaYXJFLZX24} and generation~\cite{LLMCodeSearch24, DBLP:conf/icse/IzadiKDOPD24, DBLP:conf/icse/Du0WWL0FS0L24}. Thus, LLMs have the potential to understand contextual information and generate code based on such information. 
Our insight is to leverage the code understanding ability of LLMs to mine the intention of MR and input relation from the hard-coded test inputs and SUT's function, and take advantage of their code generation ability to produce good input transformation code.
We propose three designs to harness LLMs' abilities.

\footnotetext{This {MR-encoded test case} is crafted from \code{org.hisp.dhis.util} in project \textsc{dhis2-core}, where 
long format date is ``\code{yyyy-mm-dd hh:mm:ss}'' and medium format date is ``\code{yyyy-mm-dd}''.}

\textit{\textbf{Firstly}}, we observe that directly providing LLMs with contextual information only results in \todo{around 50\% generalizable transformations (Section~\ref{sec:rq-toolEffectivenss})}. This is unsatisfactory. 
To address this, we need a design that allows LLMs to effectively express the input relation inferred from the hard-coded inputs and generate transformation code. 
\textit{To realize this goal,} we design \tool with two phases.
In \phaseOneIndex, LLMs perform \todo{analogical reasoning \cite{ICLR24_Analogical, analogical23nature}} on the hardcoded source-followup input pairs to infer new input pairs that obey the same input relation.
In \phaseTwoIndex, LLMs generate an input transformation function based on (i) the input pair hard-coded by developers and (ii) additional input pairs generated by LLMs in \phaseOneIndex.
This design not only enables LLMs to generate code in \todo{their} familiar setup (where a task description and several examples are provided)~\cite{alphacode}, but also mitigates the {above-mentioned} overfitting issue due to the limited number of examples. 

\textit{\textbf{Secondly}}, we found that LLMs \todo{often} generate task-irrelevant code segments, of which some are even faulty. 
For example, when tasked with generating a test input, an LLM might include an incorrect assertion statement.
\tool addresses this by refining the LLM-generated code through data-flow analysis, extracting only the relevant code for the given task (i.e., additional input pairs and input transformation generation).

\textit{\textbf{Thirdly}}, to mitigate the errors in the relevant codes generated by LLMs, \tool leverages the developer-written output relations (i.e., assertions) {in MTCs} as oracles to verify the generated test pairs. We further employ additional inputs to select the most generalizable input transformation as the result.

We evaluated \tool with \todo{100} developer-written test cases that encode MRs. 
The results show that \tool can generate compilable input transformations for \todo{95} MRs, where \todo{72} can generalize to all potential source inputs prepared in our evaluation. 
\tool generates 17.28\% more compilable transformations and 33.33\% more generalizable transformations than directly prompting \gpt. 
Besides, \tool-generated transformations produce follow-up inputs for 91.21\% source inputs, representing a 122.10\% improvement over \gpt{} in generating follow-up inputs. 
Our ablation study indicates that all three designs (i.e., additional input pairs, date-flow analysis based refinement, and output-relation based validation) contribute to \tool's overall performance, with validation and additional input pairs having the most impact. 
Furthermore, incorporating MRs with input transformations and new source inputs leads to 10.62\% and 18.91\%  increases in line coverage and mutation score on top of developer-written test cases, demonstrating the practical usefulness of \tool-generated transformations in enhancing test adequacy.


\textbf{Contribution.} Our work makes the following contribution:
\begin{itemize}[leftmargin=*]
    \item To the best of our knowledge, we are the first to generate input transformations for MRs encoded in test cases. With the generated input transformations, more encoded MRs can be reused to enhance the test adequacy of SUTs. 
    \item We propose \tool{}, an LLM-based approach to deduce input transformation functions. By generating multiple example input pairs, \tool mitigates overfitting and produces generalizable transformations. It also incorporates a code refinement strategy based on data-flow analysis and a validation strategy to mitigate the faulty irrelevant code generated by LLMs. This design can be applied to other code generation tasks.
    \item We extensively evaluate \tool's effectiveness in generating input transformations.
    Results show that \tool can generate effective input transformations, where \todo{72\%} input transformations are generalizable to all prepared source inputs. 
    When integrated with these transformations, the encoded MRs increase line coverage by \todo{10.62\%} and mutation score by \todo{18.91\%}.
    \item We build a dataset of 100 encoded MRs dated after {01-April, 2023}, and released it with our replication package on the website~\cite{tool}.
\end{itemize}
\section{Preliminaries} 

\subsection{Metamorphic Testing}\label{sec:preliminary-MTdef}

Metamorphic Testing (MT) \todo{validates} a program $P$ using Metamorphic Relations (MRs). 
An MR $\relation$ can be expressed as a logical implication from an \textit{\textbf{input relation}} $\inputRelation$  to an \textit{\textbf{output relation}} $\outputRelation$~\cite{mrscout, 2016-segura-tse, chen2018metamorphic}.
\[ \relation=\langle\relation_i, \relation_o\rangle, \textrm{ where }\relation_i \left( \sourceInput, \followUpInput \right) \implies \outputRelation \left( \sourceOutput, \followUpOutput \right) \]
$\inputRelation$ defines the rule to generate an additional test input (known as \textit{follow-up input} $\followUpInput$) from a given test input (known as \textit{source input} $\sourceInput$), and $\outputRelation$ defines the relation between the expected outputs ($\sourceOutput$, $\followUpOutput$) for the source and follow-up inputs, respectively.
For example, given a program \textit{$P$} implementing the \textit{sine} function, an MR can be defined over an input relation $\inputRelation$ as $x_f {=} {-}x_s$ ($\forall x_s \in \mathbb{R}$) and an output relation $\outputRelation$ as $y_f {=} {-}y_s$, based on the property that $P(x) {=} {-}P({-}x)$ should hold for a correctly implemented \textit{sine} function.

Given an MR $\relation$ for a SUT $P$, 
conducting MT for $P$ entails the following five steps: (i) constructing a source input $x_s$, (ii) executing $P$ on $x_s$ and obtaining the source output $y_s$, (iii) constructing a follow-up input $x_f$ that satisfies $\relation_i$, (iv) executing $P$ on $x_f$ and obtaining the follow-up output $y_f$, and (v) verifying if the two outputs $y_s$ and $y_f$ satisfy the output relation $\relation_o$.
Typically, a function referred to as the \textit{\textbf{input transformation}} is designed to implement $\inputRelation$ to generate $x_f$ from the given $x_s$, and $x_s$ can be written by developers or automatically generated (e.g., random testing)~\cite{mrscout, 2016-segura-tse}.



\subsection{MR-Encoded Test Cases}\label{sec:preliminary-MTC}
MR-encoded test cases (MTCs), introduced by \citet{mrscout}, are test cases encoded domain-specific knowledge that suggests useful MRs.
These MTCs are prevalent, with over 11,000 identified across 701 open-source projects in their study.
\editedA{
An MTC can be considered as an instance of an MR, already implemented with {specific source and follow-up input values}, invocations of methods under test, and output relation assertions. These output relations are well-coded and serve as oracles, requiring no further refinement to apply to new test inputs.}
Such encoded MRs can be generalized to new inputs and facilitate automated MT by incorporating automatic input generation techniques.

Consider the example in Figure~\ref{fig:overview_of_IT4MT}.
The encoded MR in this test case is: ``IF a date $x_1$ in long format (``yyyy-mm-dd hh:mm:ss'') is one day ahead of another long-format date $x_2$ ($\inputRelation$), THEN $x_1$ in medium format (``yyyy-mm-dd'') should also be one day ahead of medium-format $x_2$ ($\outputRelation$)''.
The SUT method \code{toMediumDate} is executed on the source input \code{dateA} and the follow-up input \code{dateB} separately, and the corresponding outputs are \todo{verified} by \code{assertThat(mediumB,} \code{is(plusOneDay(mediumA))}, which implements $\outputRelation$.

Such an implemented MR instance can be reused and generalized to many new inputs. 
However, the follow-up input \code{dateB} is hardcoded as \code{"2024-01-02 00:00:00"} instead of being generated from \code{dateA} by an input transformation program. 
While the $\outputRelation$ is explicitly coded, the $\inputRelation$ remains \textbf{implicit}, hidden within the specific source and follow-up input values \code{dateA} and \code{dateB}.
According to Xu et al.'s study, over 70\% of MR-encoded test cases lack explicitly coded $\inputRelation$ (i.e., input transformations).
This limitation prevents these MRs from being directly applied to new inputs automatically generated by existing tools, e.g., Evosuite~\cite{evosuite2011} and Randoop~\cite{randoop}. 
While these tools are proficient in generating diverse source inputs, they cannot generate input pairs that satisfy an input relation.

In this paper, we address this limitation by deriving an explicit input relation from a given test case and its hardcoded input pairs. Specifically, our goal is to construct an \textbf{input transformation function} that converts a source input into a follow-up input, as shown in Figure~\ref{fig:IT4MTb}.
With such input transformations, embedded MRs can be reused with a broader range of test inputs (Figure~\ref{fig:IT4MTc}) to exercise more SUT's behaviors, thereby enhancing test adequacy.
Additionally, these developer-written MRs serve as reliable oracles for new test generation.

\begin{figure*}[t]
\centering
\includegraphics[width=.8\textwidth]{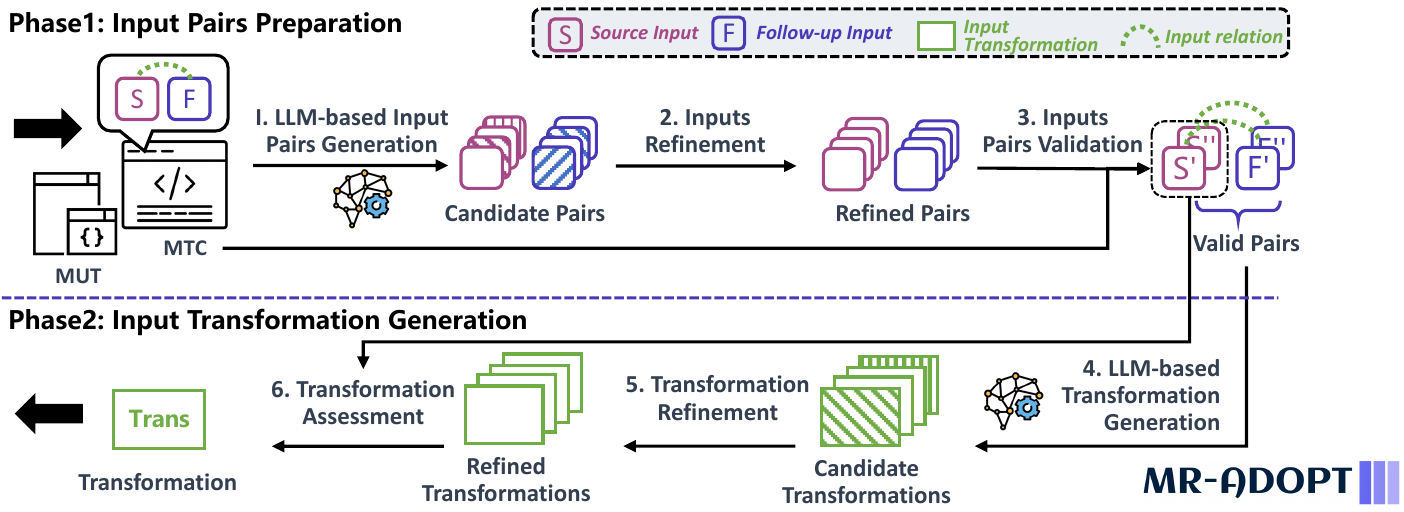}
\vspace{-10pt}
\caption{An overview of \tool.} \label{fig:overview}
\end{figure*}

\section{\tool} 

Figure~\ref{fig:overview} presents an overview of \tool.
It takes {a pair of source and follow-up inputs}, along with its context (i.e., an MR-encoded test case (MTC) and methods under test (MUT)), and outputs an input transformation function.\jr{I think we should explain how we identify the source and follow-up inputs and the input-transformation function. This is nontrivial}\hc{I think the source and follow-up inputs are identified as inputs to \tool{}.}
\tool works in a two-phase pipeline. In the first phase, it generates additional source-follow-up input pairs and uses them as examples to better describe the input relation, which provides useful guidance for the generation of input transformations. 
In the second phase, it generates input transformation functions based on these example pairs.
This setup, familiar to LLMs for code generation tasks, includes a task description and several examples~\cite{alphacode}, providing more information to effectively guide LLMs in generating generalized transformations.

In each phase, \tool employs generation, refinement, and validation procedures.
In \phaseOneIndex{}, \tool first leverages LLMs to generate candidate test input pairs, then refines them based on data-flow analysis to exclude irrelevant code that can contain errors, and finally filters valid input pairs based on output relation assertions.
In \phaseTwoIndex{}, \tool leverages LLMs to generate candidate input transformations based on the input pairs from \phaseOneIndex.
These candidate transformations are then refined by removing irrelevant code elements and adding dependencies, and assessed by applying them to additional source inputs.
Ultimately, \tool outputs the most generalizable transformation function.


\subsection{\phaseOneLong}\label{sec:app-phase-inputsGen}

\subsubsection{Input Pair Generation}\label{sec:app-phase1-inputGen}

\tool uses an LLM to produce new source-followup input pairs by imitating a given input pair within the context of an existing MTC (which includes the input pair and developer-written assertions checking the output relation) and corresponding methods under test. 

\begin{figure}
\centering
\includegraphics[width=.5\textwidth]{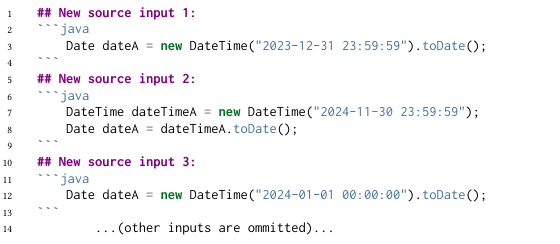}
\caption{Examples of LLM-generated source inputs}
\label{lst:LLM-generated-Sinput}
\end{figure}

Following the idea of the Chain of Thought strategy~\cite{DBLP:conf/nips/Wei0SBIXCLZ22}, \tool prompts an LLM in two steps: first to generate source inputs, and then to generate the corresponding follow-up inputs.
This step-by-step approach is adopted because we found that LLMs perform better when generating source and follow-up inputs sequentially rather than generating entire input pairs at once. 
Our source input generation prompt follows recent practices~\cite{arxiv23_FDUGPT4TestGen, DBLP:journals/corr/abs-2404-14646}, and includes (i) a system message about the role of a Java expert and the task to generate test inputs, (ii) the code of methods under tests (MUTs), (iii) the code of the MR-encoded test case (MTC), and (iv) the output format.
Such a prompt provides necessary contextual information (ii and iii) and task description (i and iv) for generating source inputs. 
Detailed prompt templates and examples are available on \tool's website~\cite{tool}. 
\jr{I think we can at least give a very short skeleton}
Figure~\ref{lst:LLM-generated-Sinput} shows several example source inputs generated by \todo{GPT-3.5} with this prompt.

\begin{figure}
    \centering
    \includegraphics[width=.5\textwidth]{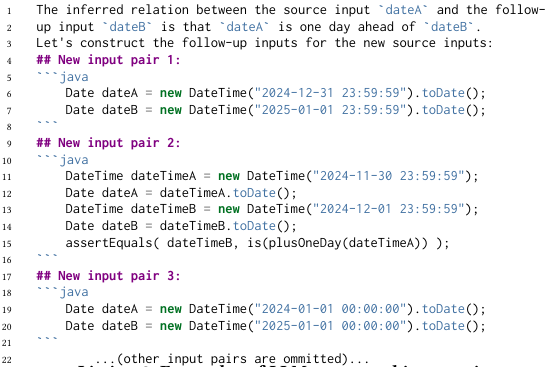}
    \caption{Examples of LLM-generated input pairs}
    \label{lst:LLM-generated-inputPair}
\end{figure}

The follow-up input generation prompt is similar to the source input prompt, with the key difference being the addition of previously generated example source inputs to guide the creation of follow-up inputs. We also adjust the task description and output format to instruct LLMs to generate source-followup input pairs, using the original pair in the MTC as a sample.
\editedA{
There is a trade-off between providing enough examples to guide the following generation of generalizable transformations and maintaining efficiency in terms of time and cost. To balance these factors, we use five examples.
Following recent studies' nucleus sampling~\cite{DBLP:conf/iclr/HoltzmanBDFC20, DBLP:conf/icse/Du0WWL0FS0L24}, \tool repeats the above generation process five times with a temperature setting of 0.2~\cite{DBLP:journals/corr/abs-2403-16898, DBLP:journals/corr/abs-2107-03374}.}
\todo{Figure~\ref{lst:LLM-generated-inputPair} shows several example input pairs.}


\subsubsection{Input Pair Refinement}\label{sec:app-phase1-inputRefine}

Consistent with observations in existing studies~\cite{DBLP:journals/corr/abs-2401-01701}, we found that LLM-generated codes often include irrelevant segments, which may introduce errors.
For instance, the second input pair in Figure~\ref{lst:LLM-generated-inputPair} includes an assertion statement \code{assertEquals(dateTimeB, is(plusOneDay(dateTimeA)));} alongside the desired input pair code.
This assertion is irrelevant to constructing \code{dateA} or \code{dateB} and incorrectly calls the method \code{plusOneDay(Date date)} with a \code{DateTime} object, resulting in a type mismatch exception. 

Our task focuses on constructing source and follow-up inputs.
\editedA{
To exclude such irrelevant code and bypass unnecessary errors, we perform a data-flow analysis on the LLM-generated code and build a dependency graph. 
\tool constructs data dependency graphs for source/follow-up inputs by first identifying statements that define source/follow-up inputs, extracting variables used for these definitions, and then recursively tracing these variables’ dependencies. 
Finally, \tool isolates the dependent statements related to the source and follow-up inputs and removes other statements.
}
For example, in the second input pair of Figure~\ref{lst:LLM-generated-inputPair}, the source input \code{dateA} and follow-up input \code{dateB} depend on objects \code{dateTimeA} and \code{dateTimeB}, respectively. Thus, the statements (Lines \todo{11-14}) for constructing  \code{dateA}, \code{dateTimeA}, \code{dateB}, and \code{dateTimeB} are considered relevant, while the assertion statement (Line 15) is excluded.
Finally, \tool retains only the statements relevant to constructing source and follow-up inputs, excluding all other irrelevant statements from the LLM-generated code.


\subsubsection{Input Pair Validation}\label{sec:app-phase1-inputValidate}
The previous refinement step removes the irrelevant code segments generated by LLMs and results in candidate source-followup input pairs. 
However, a pair of inputs still can be \textit{invalid} if they violate the input relation of an encoded MR. 
For example, the third input pair shown in Figure~\ref{lst:LLM-generated-inputPair} is an invalid test pair.
The input relation of the embedded MR is that ``\code{dateA} is one day ahead of \code{dateB}'', 
while an LLM generates an input pair of ``2024-01-01'' and ``2025-01-01''.
Such test case pairs do not align with the intended input transformation and can mislead the generation of transformations.
To address this, we propose a method to discard such invalid input pairs.

\begin{figure}
    \centering
    \includegraphics[width=.5\textwidth]{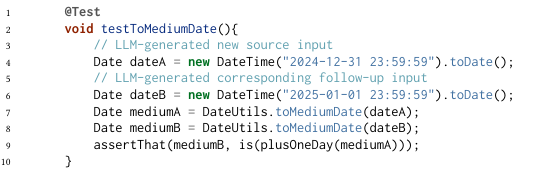}
    \caption{Validating an LLM-generated input pair}
    \label{lst:validate-inputPair}
\end{figure}

We use the output relation of an encoded MR to validate LLM-generated input pairs.
Specifically, \tool executes SUT on generated input pairs and checks the outputs against the output relation, which is an explicit reusable code in the MTC, i.e., the developer-written assertions (Line 7 in Figure~\ref{fig:IT4MTa}). 
For each input pair, if its outputs of invoking methods under test on the inputs pass the developer-written assertions, \tool considers it a valid input pair.
As shown in Figure~\ref{lst:validate-inputPair}, 
if the outputs \code{mediumA} and \code{mediumB} pass the assertion (Line~\todo{9}), the inputs \code{dateA} and \code{dateB} are considered valid.
\todo{This step aims to filter out invalid input pairs generated by LLMs from the example set. It could discard some source-followup input pairs that match the input relation in fact. Factors such as the bugs in a non-regression SUT may lead to false violations and mistaken deletions of these pairs. However, the goal of the first phase is to prepare examples that give more information about the input relation for the second phase. Thus, it does not require \textit{complete} source-followup input pairs.}


\subsection{\phaseTwoLong}\label{sec:app-phase-transGen}

\subsubsection{Transformation Generation}\label{sec:app-phase2-transGen}


In this step, \tool instructs an LLM to generate candidate input transformation functions for an encoded MR by providing example source-followup input pairs. The examples include the original hard-coded pair and additional pairs generated in~\phaseOneIndex.
The prompt for transformation generation is similar to the input pair generation prompt~(Section~\ref{sec:app-phase1-inputGen}), consisting of (i) a system message, (ii) the code of MUT, (iii) example input pairs, (iv) the code of an MTC, and (v) the output format. 
The difference is that the task shifts from generating source-followup input pairs to generating input transformation functions, whose parameter list and return type are already specified. 
Detailed prompt template and samples are available on \tool's website~\cite{tool}.



\begin{figure}
    \centering
    \includegraphics[width=.5\textwidth]{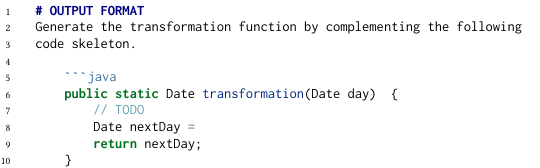}
    \caption{An example of output format in the prompt}
\label{lst:transSkeleton}
\end{figure}


\todo{Figure~\ref{lst:transSkeleton}} shows the output format specified in the prompt, which defines the skeleton of the input transformation function to generate.
It includes the function name, parameter (i.e., source input) types and names, and type of the return value (i.e., follow-up input)~\footnote{
For MRs with multiple follow-up inputs, the return type is a list of objects.
}.
Following recent studies' nucleus sampling~\cite{DBLP:conf/iclr/HoltzmanBDFC20, DBLP:conf/icse/Du0WWL0FS0L24}, for each MR, \tool instructs an LLM to generate one input transformation function, and repeats the generation process five times with a temperature setting of 0.2~\cite{DBLP:journals/corr/abs-2403-16898, DBLP:journals/corr/abs-2107-03374}. Finally, five candidate transformation functions can be generated.

\tool extracts the generated functions by identifying code blocks wrapped with \verb|```| and extracting the code that matches the given transformation function skeleton. 
This ensures the generated code conforms to the required format and can be easily integrated into given MR-encoded test cases. 

\subsubsection{Transformation Refinement}\label{sec:app-phase2-transRefine}
\begin{figure}
    \centering
    \includegraphics[width=.5\textwidth]{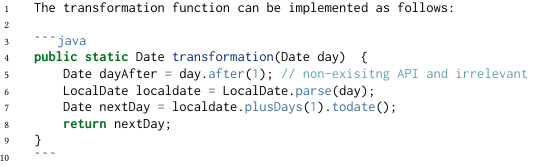}
    \caption{An example of LLM-generated transformation}
    \label{lst:LLM-generated-tran}
\end{figure}

Similar to the situation discussed in Section~\ref{sec:app-phase1-inputRefine}, LLM-generated transformation functions can contain irrelevant code, some of which can cause errors (e.g., invoking non-existing APIs). 
\editedA{
To address this issue, \tool constructs data dependency graphs for follow-up inputs by first identifying statements that define these inputs. It then extracts the variables used in these definitions and recursively traces their dependencies. 
Statements involved in the dependency graph are considered relevant and retained, while others are considered irrelevant and excluded.}
As the example shown in Figure~\ref{lst:LLM-generated-tran}, the follow-up input \code{nextDay} depends on \code{localDate}, which further depends on \code{day}. Statements constructing \code{nextDay} and \code{localDate} are retained, while irrelevant statements such as \code{Date dayAfter=day.after(1)} are excluded.



\editedA{
After excluding irrelevant code, \tool analyzes and imports dependencies needed by the generated transformation function. 
Using JavaParser~\footnote{https://javaparser.org/}, \tool identifies dependent class names through syntax analysis.
It then retrieves potential classes defined or imported in {source} and imports those whose names match the dependent classes.
For example, the internal class \code{LocalDate} Figure~\ref{lst:LLM-generated-tran} will be imported. 
Although matching dependent classes by name can be inaccurate if LLM-generated code uses incorrect names or third-party libraries not imported in the project, our experimental observations indicate that this issue is minor. Most dependent classes are derived from the context provided to LLMs, such as the MUT or MTC code.
}



\subsubsection{Transformation Assessment}\label{sec:app-phase2-transValid}

After refining candidate transformations, \tool further assesses their quality by applying them to new source inputs.
In this step, \tool leverages new source inputs generated in Section~\ref{sec:app-phase-inputsGen} to assess the generalizability of candidates and then selects the most generalizable one.

\begin{figure}
    \centering
    \includegraphics[width=.5\textwidth]{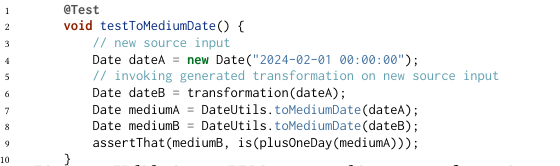}
    \caption{Validating an LLM-generated input transformation with a new source input}
    \label{lst:validate-trans}
\end{figure}

\begin{figure}
    \centering
    \includegraphics[width=.5\textwidth]{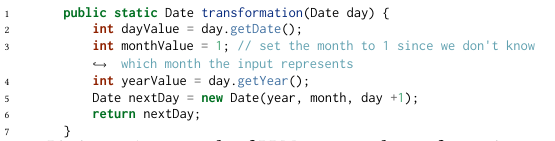}
    \caption{An example of LLM-generated transformation}
\label{lst:LLM-generated-tranNonGeneral}
\end{figure}

Specifically, \tool uses new source inputs as test inputs and employs developer-written assertions (i.e., output relation assertions) as test oracles. A transformation is considered applicable to a given source input if (a) the input transformation function can successfully generate a corresponding follow-up input without throwing exceptions, and (b) the outputs from executing the methods under test pass the developer-written assertions.
For example, in Figure~\ref{lst:validate-trans}, given the source input \code{dateA}, if the follow-up input \code{dateB} is successfully generated and the outputs \code{mediumA} and \code{mediumB} pass the assertion (Line~\todo{9}), \tool considers the transformation applicable to input \code{dateA}. Conversely, Figure~\ref{lst:LLM-generated-tranNonGeneral} shows a failing transformation that only works for January dates.
\tool assesses all candidate transformation functions using both new and the original source input. It then selects the most generalizable transformation that applies to the most inputs. 
In case of a tie, \tool returns the first generated one as the result.

\section{Evaluation}\label{sec:evaluation}
\jr{I noticed a mixed use of tenses. Fix to past or present}

\subsection{Research Questions}\label{sec:RQdefs}

Our evaluation aims to answer the following research questions: 

\begin{itemize}[leftmargin=*]
\item \textbf{RQ1:} \textit{How effective is \tool in generating input transformations?}
This RQ compares the quality of the input transformation functions generated by \tool and baselines to evaluate the effectiveness of \tool in generating generalizable input transformations for MRs encoded in test cases.

\item \textbf{RQ2:} \textit{How effective are \tool-generated input transformations in constructing follow-up inputs, compared with LLMs?} 
This RQ investigates the benefits of generating input transformation functions, by comparing the quality of follow-up inputs produced by these functions versus those directly generated by LLMs.

\item \textbf{RQ3:} \textit{What is the contribution of each component in \tool?}
This RQ performs an ablation study to reveal how each component contributes to generating input transformations. 

\item \textbf{RQ4:} 
\textit{How useful are encoded MRs in enhancing test adequacy with the generated input transformations?}
With input transformations generated by \tool, more encoded MRs can be reused with new inputs to test more behaviors of SUT.
This RQ investigates the usefulness of such encoded MRs in improving test adequacy, demonstrating the usefulness of generating input transformation.
\end{itemize}

\subsection{Dataset} \label{sec:dataset}

\textbf{MR-encoded test cases (MTCs).} 
We followed \citet{mrscout} to collect high-quality Java projects with at least 200 stars over GitHub. 
\editedA{
Besides, we further excluded the projects created before 01-April 2023 to prevent the experimental LLMs from having potentially learned the code during training, thereby reducing the potential for data leakage~\cite{DBLP:journals/corr/abs-2403-16898}.}
Finally, we collected 2,007 MTCs from qualified projects.
From these MTCs, we retained test cases that (i) can be successfully compiled, (ii) can be successfully executed (i.e., passing developer-written assertions), and (iii) contain MRs associated with exactly two method invocations (one for the source input and one for the follow-up input). 
\editedA{
The third criterion excludes complex and less common MRs involving multiple input groups~\cite{mrscout}. MRs with exactly two invocations constitute the majority (65\%) of MTCs~\cite{mrscout}. Handling MRs with more than two invocations presents significant challenges in identifying source and follow-up inputs. This problem is nontrivial and remains an important future work.}
Finally, we obtained 180 MTCs, including 54 with explicit input transformations written by developers and 126 without such transformations, consistent with the distribution reported by~\citet{mrscout}. 

\vspace{0.02cm}
\noindent \textbf{Generation Tasks and Ground Truths.} 
Based on the collected 180 MTCs, we prepared a dataset containing (i) 100 MTCs without input transformations as tasks, and (ii) corresponding input transformation functions as ground truths. The preparation process is as follows.
Firstly, we tried to utilize all 54 MTCs with ground truths, i.e., developer-written input transformations.
For each MTC, we executed the input transformation on the hardcoded source input to obtain the follow-up input. 
We prepared a task by replacing the developer-written transformation with the hardcoded follow-up input.
Some MTCs whose follow-up input cannot be hardcoded are excluded.
For example, an MR for a text render class is ``the width of a text (source input) should not be greater than its bold version (follow-up input)''. The follow-up input (bold text) can only be generated by a method \code{bold()}, which is a developer-used transformation program.
Finally, we built \todo{36} tasks from 36 MTCs with developer-written transformations. 

Next, we manually constructed input transformation functions for MTCs lacking developer-written transformations. Specifically, 64 out of 126 MTCs without input transformations were randomly selected as tasks.
For each task, one author of this paper examined the SUT and its underlying MRs and then created a transformation function applicable to the original source input and generalizable to new valid source inputs. 
Another author reviewed these transformations, and any disagreements were discussed and resolved with consensus.
This process took approximately 200 human hours.
Details of this dataset can be found on \tool's website~\cite{tool}.

\subsection{Environment and Large Language Models} \label{sec:eval-environment-LLMs}
Our experiments were conducted on machines with three RTX4090 GPUs, dual Intel Xeon E5-2683 v4 CPUs, and 256 GB RAM.

The large language models used in our evaluation include GPT-3.5 from OpenAI~\footnote{https://platform.openai.com/docs/models/} and three open-source code models: Llama3-8B~\footnote{https://huggingface.co/meta-llama/Meta-Llama-3-8B-Instruct} from Meta, Deepseek-coder-7b~\footnote{https://huggingface.co/deepseek-ai/deepseek-coder-7b-instruct-v1.5} from DeepSeek, and CodeQwen1.5-7B-Chat~\footnote{https://qwenlm.github.io/blog/codeqwen1.5/} from Alibaba. 
We use these LLMs since they are popular state-of-the-art~\footnote{https://evalplus.github.io/leaderboard.html} code models in well-known LLM families and deployable at our machines. 


\subsection{Source Input Preparation} \label{sec:experimental-source-inputs}

To evaluate the generated input transformations, we need new valid source inputs as a ``test set''.
As introduced in Section~\ref{sec:preliminary-MTdef}, automatic test input generation techniques (such as Evosuite~\cite{evosuite2011} or Randoop~\cite{randoop}) can be employed to prepare source inputs. 
However, we found that these tools often fail to generate test inputs for many MRs.
This is because over 50\% of experimental MRs' inputs are user-defined complex objects with complicated preconditions and environments, which are challenging for tools like Evosuite to handle. This aligns with Xu et al.'s observation~\cite{mrscout}. 

Recent studies show that LLMs are good test input generators \todo{\cite{arxiv23_FDUGPT4TestGen, DBLP:journals/corr/abs-2404-14646}}.
In this study, we employed an LLM (\qw{}) to generate new source inputs for evaluating transformations, while other experimental LLMs were used to generate transformations.
As a reminder, to mitigate circular evaluation, we employed different LLMs for preparing the ``test set'' and for generating input pairs and transformation functions in \tool.
We reused the prompt template from \tool's \phaseOneIndex. \qw{} was instructed to generate five source inputs at a time, and the process was repeated ten times with a 0.2 temperature setting to produce more source inputs.

For the 100 experimental MRs, \qw{} generated a total of 5,355 new source inputs.
We first filtered out 3,058 duplicate inputs using string matching. 
Next, we identified valid source inputs by executing them on the corresponding ground truth transformations.
A source input is considered valid only if the ground truth transformation successfully generates a follow-up input, and the outputs of this source input and corresponding follow-up input pass the developer-written assertions (\todo{$R_o$}).
\qw{} failed to generate a new valid source input for 5 MRs whose inputs are complex objects and have strict domain-specific constraints.
Finally, we collected 1,366 valid source inputs, averaging 14.37 per MR.

\begin{table}[]
\footnotesize
    \caption{Effectiveness of \tool in generating input transformations for 100 MRs encoded in test cases}\label{tab:rq-effectivess-tool}
    \centering
    \vspace{-10pt}



\newcommand{\improve}[1]{\scriptsize{(+#1\%)}} 

\setlength{\tabcolsep}{1.6pt}
\centering
\begin{tabularx}{\linewidth}{l|ccc|ccc}
\toprule
\multirow{3}{*}{\textbf{Metric (\# Trans.)}} &
\multicolumn{3}{c|}{\textbf{ Direct Prompting  }} & 
\multicolumn{3}{c}{\textbf{ \tool }} \\
\cmidrule(l{1pt}r{1pt}){2-4} 
\cmidrule(l{1pt}r{1pt}){5-7} 
& \textit{\lm} & \textit{\dps } & \textit{ \gpt }
& \textit{\lm} & \textit{\dps } & \textit{ \gpt }\\

\midrule
compilable                      & 79     & 80     & 81     & 86 \improve{8.86}     & 89 \improve{11.25}    & \textbf{95 \improve{17.28}}     \\ 
\textgreater{}0\% generalizable   & 69     & 72     & 69     & 77 \improve{11.59}    & 82 \improve{13.89}    & \textbf{83 \improve{20.29}} \\ 
\textgreater{}75\% generalizable  & 64     & 67     & 63     & 74 \improve{15.66}    & 80 \improve{19.40}    & \textbf{81 \improve{28.57}} \\ 
100\% generalizable               & 57     & 60     & 54     & 68 \improve{19.30}    & 71 \improve{18.33}    & \textbf{72 \improve{33.33}} \\ 
\bottomrule
\end{tabularx}
\begin{tabularx}{\linewidth}{l}
\textit{\# $n$\% generalizable:} the number of generated input transformations applicable to at \\least $n$\% of source inputs. 
\end{tabularx}


    \vspace{-10pt}
\end{table} 


\subsection{RQ1: Effectiveness of \tool} \label{sec:rq-toolEffectivenss}
\subsubsection{Experiment Setup.}
This RQ inspects \tool's effectiveness in generating input transformation functions by examining their comparability and generalizability to new source inputs. 

\textbf{Baselines.} \label{sec:baseline-effectGenTrans}
To the best of our knowledge, no existing approach generates input transformation functions for MRs across different domains.
Given the proven effectiveness of LLMs in code and test generation, we set \textit{directly prompting LLMs} as a baseline.
Specifically, we directly prompted \gptlong{}, \lmlong{}, and \dpslong{} (shorten as \gpt, \lm, and \dps, respectively). 
\editedA{
The template is similar to \tool's and available at~\cite{tool}.}
The knowledge cut-off dates for these models are September 2021 \footnote{https://help.openai.com/en/articles/8555514-gpt-3-5-turbo-updates}, March 2023 \footnote{https://huggingface.co/NotAiLOL/Meta-Llama-3-8B-Instruct}, and March 2023 \footnote{https://github.com/deepseek-ai/DeepSeek-Coder/issues/89}, respectively, before the creation date of our dataset's MTCs~(Section~\ref{sec:dataset}), 


\jr{Is the comparison with the baseline fair? We may need to justify this because no (even simple) post-processing seems unfair to the baselines.}

\textbf{Configuration of Baseline LLMs.}
\label{sec:LLM-configuration}
Following recent studies~\cite{DBLP:conf/icse/Du0WWL0FS0L24}, we used the nucleus sampling~\cite{DBLP:conf/iclr/HoltzmanBDFC20} and repeated the generation process five times for each task with a temperature setting of 0.2~\cite{DBLP:journals/corr/abs-2403-16898, DBLP:journals/corr/abs-2107-03374}, and selected the best result for comparison. 
The configuration of \tool was introduced in Section~\ref{sec:app-phase2-transGen}.



\textbf{Metrics.} For this RQ, we introduced two metrics: (i) \textit{\# compilable transformations:} the number of generated input transformations that can successfully compile, and (ii) \textit{\# $n$\% generalizable transformations:} the number of generated input transformations applicable to at least $n$\% of source inputs prepared in Section~\ref{sec:experimental-source-inputs} ($n=0, 75, 100$ representing at least one, upper-quartile, and all inputs, respectively).
A transformation $t$ is considered \textit{applicable} to a source input $x_s$ if $t$ generates a follow-up input $x_f$ for $x_s$, so that a \textit{correct} SUT does not violate the output relation on the input pair ${<}x_s, x_f{>}$.

\subsubsection{Result.}
As shown in Table~\ref{tab:rq-effectivess-tool}, \tool effectively produced many compilable input transformation functions that well generalize to prepared source inputs. We found that \tool works best with \gpt{}. Specifically, using \gpt{} (the last column), \tool produced compilable transformations for 95 out of 100 MRs, with 72 of these transformations effectively applied to \textit{all} prepared source inputs.
\tool also works well with \lm{} and \dps{}, generating 68 and 71 (100\%) generalizable transformations, respectively. 
\jr{I cannot understand what this paragraph tries to do. It seems quite messy.} 
Besides, we found that some generated transformations generalize well to some, but not all, source inputs prepared in our experiment. Specifically, with \gpt{}, 83 out of 95 compilable transformations applied to at least one source input, and 81 of them applied to more than 75\% of the prepared source inputs. Similar results were found with \lm{} and \dps{}.
We considered these transformations generated by \tool \textit{still useful} to some extent, as they successfully prepare some valid input pairs.
Upon further analysis, we found that their limitations could potentially be addressed with more comprehensive prompts to handle corner cases. LLM-generated transformations effectively handle common cases but struggle with edge cases. For example, an ideal transformation would generate a higher version string in any scenario (e.g., transforming \code{"1.0-A1"} to \code{"1.0-B1"}), but the LLM-generated transformation relies on a 'Major.Minor.Revision' convention (e.g., \code{"1.0.1"}) and fails with cases like \code{"1.0-A1"}.

There were 5, 14, and 11 transformations generated by \tool with \gpt{}, \lm{}, and \dps{}, respectively, that failed to compile.
The main reasons include: (i) the generated transformations invoke non-existing methods to generate the follow-up input and (ii) they invoke inaccessible APIs due to permission restrictions (e.g., private methods).
Additionally, the compilable but not generalizable transformations were primarily due to \phaseOneIndex{} failing to generate valid input pairs for these MRs, leading to LLM-generated transformations overfitted to the given input pair.

\editedA{
We also compared \tool's performance (columns 5-7) with the baseline of directly prompting LLMs (columns 2-4). 
Although the output relation encoded in MTC was provided in prompts for baselines to aid transformation generation, \tool still generated more compilable transformations. This improvement is due to \tool's code refinement and assessment strategies.}
Moreover, \tool demonstrates substantial improvements in generating transformations that are $>$75\% and 100\% generalizable, with increases of 15.66\% to 28.57\% and 18.33\% to 33.33\%, respectively. 
This suggests the effectiveness of preparing more examples for LLMs and the benefits of \tool's refinement and selection strategies.

\begin{answertorq}
\tool significantly outperforms the baseline LLMs across all metrics.
 Compared to directly prompting LLMs, \tool achieves \todo{18.33\%$\sim$33.33\%} improvement in generating \todo{100\% generalizable} input transformations. 
\end{answertorq}

\subsection{RQ2:~Effectiveness~of~Input~Transformations} \label{sec:rq-transforamtionEffectivenss}
\subsubsection{Experiment Setup.}
This RQ examined the quality of follow-up inputs produced by input transformations generated by \tool.
We set LLMs as the baselines because they are off-the-shelf black-box transformations that can generate follow-up inputs given source inputs, as introduced in Section~\ref{sec:app-phase1-inputGen}.
We also included LLMs enhanced with \tool{}'s refinement procedure (marked with $^+$) for comparison. This can reflect the effectiveness of \tool's refinement for input pairs preparation (Section~\ref{sec:app-phase1-inputRefine}).


\textbf{Metric.} We generated follow-up inputs by feeding the \todo{1,366} prepared source inputs (Section~\ref{sec:experimental-source-inputs}) to input transformations generated by \tool and the vanilla LLM baselines.
To compare the qualities of the follow-up inputs produced by the \tool-generated transformations and the baselines, we used the number of {\textit{valid} follow-up inputs as the metric.
Similar to Section~\ref{sec:rq-toolEffectivenss}, we consider a follow-up input $x_f$ \textit{valid} if it and its corresponding source input can pass developer-written output relation assertions.



\begin{table}[]
\footnotesize
    \caption{Effectiveness of \tool's transformations in constructing follow-up inputs for 1366 source inputs}\label{tab:rq-effectiveness-trans} 
    \centering
    \vspace{-10pt}
\begin{tabular}{c|ccc|c}
\toprule
\textbf{\tool} & \lm{}  & \dps{}  & \gpt{}  & \textbf{Improvement}                  \\
\midrule
1246 & 697 & 724 & 597  &   +72.10\%$\sim$\textbf{+108.71\%} \\ 
\midrule
\textbf{\tool} & \lm{}$^+$ & \dps{}$^+$ & \gpt{}$^+$ &  \textbf{Improvement}                  \\
\midrule
1246 & 770 & 737 & 708  &  +61.82\%$\sim$\textbf{+75.99\%}  \\ 
\bottomrule
\end{tabular}
\begin{tabularx}{\linewidth}{l}
$^+$ means incorporating \tool{}'s input refinement procedure for LLMs' answers.
\end{tabularx}



\end{table}

\subsubsection{Result.}


As shown in Table~\ref{tab:rq-effectiveness-trans}, when built with \gpt{}, input transformation functions generated by \tool produced valid follow-up inputs for 1246 out of 1366 (91.22\%) source inputs.
The high validity rate demonstrated that \tool contributed to abundant useful source-followup input pairs.
\congying{we may not keep the sentence. (commented one)}

In comparison, three vanilla LLMs only generated valid follow-up inputs for 697 (51.02\%), 724 (53.00\%), and 597 (43.70\%) source inputs, respectively. \tool surpassed them by 72.10\%-108.71\%.
LLMs enhanced with \tool's \todo{input} refinement procedure (marked with $^+$) worked better than the vanilla LLMs.
This indicates the usefulness of our design to refine the LLM-generated \todo{test inputs (Section~\ref{sec:app-phase1-inputRefine})}.
Meanwhile, \tool's transformations still outperformed the enhanced LLMs by generating 61.82\% more valid follow-up inputs than \lm{}$^+$, 69.06\% more than \dps{}$^+$, and 75.99\% more than \gpt{}$^+$. 
This significant performance gap highlights the effectiveness of \tool's transformation functions compared to the state-of-the-art LLMs.
It also evidenced the usefulness of our idea to codify the input transformation by leveraging the code understanding and generation abilities through the two-phase pipeline and preparation-refinement-validation process.

We also summarized {two} major limitations of using vanilla LLMs as black-box transformations based on our observation. 
Firstly, LLMs can generate a follow-up input with a wrong value, which is similar to the case in Figure~\ref{lst:LLM-generated-inputPair}. 
Another limitation is that LLMs often fail to capture the constraints between multiple arguments of the follow-up input. For instance, consider a method \code{deserial(data, size)} to deserialize an \code{ArrayList} \code{data} with a given \code{size}. The \code{size} should not be greater than the length of \code{data}. However, LLMs may miss this constraint and generate invalid value for \code{size}.
These issues about value processing could be due to LLMs' limited inference ability. Instead, \tool asks LLMs to codify the input transformation and uses the code to do calculation and processing, which is recognized as a better way to exert LLMs' abilities~\cite{ChainOfCode}. 
Besides, we argued that using LLMs as transformations can be costly since we need to request LLMs for each source input. Meanwhile, \tool uses LLMs to generate transformations for once, and there is no need to query LLMs when using the generated transformations.
\congying{may add to Discussion section.}

\begin{answertorq}
\tool's refinement step can effectively enhance follow-up input generation, with up to 18.59\% improvement for \gpt{}. 
Additionally, \tool-generated transformations can effectively generate follow-up inputs for 91.21\% source inputs, surpassing \gpt{}$^+$ by 75.99\%.
\end{answertorq}

\subsection{RQ3: Ablation Study on \tool}\label{sec:rq-ablation}
\subsubsection{Experiment Setup.}
We created \todo{three} variants $v_1$, $v_2$, and $v_3$ of \tool by ablating three components to analyze the helpfulness of these designs for generating \todo{generalizable} input transformations. 
\todo{We chose \tool built with GPT-3.5 which achieves the best result in RQ1 (Section~\ref{sec:rq-toolEffectivenss}).}
The variants are as follows:
\begin{itemize}[leftmargin=*]
\item \textbf{$v_1$: \tool w/o additional input pairs.}
This variant used only one source-followup input pair hard-coded in an MTC to guide the input transformation generation. It did not use additional input pairs generated in \tool's \phaseOneIndex{} (Section~\ref{sec:app-phase-inputsGen}). 
\item  \textbf{$v_2$: \tool w/o refinement step.} This variant disabled the refinement step for generated input transformations in \tool (Section~\ref{sec:app-phase2-transRefine}). 
\item  \textbf{$v_3$: \tool w/o assessment step.} This variant disabled the assessment step for selecting the most generalizable transformations (Section~\ref{sec:app-phase2-transValid}). Instead, it randomly selected one of the compilable transformation functions as the result. 
\end{itemize}


\begin{table}[h!!]
\footnotesize
    \caption{Contribution of each component in \tool}\label{tab:rq-ablation}
    \centering
    \vspace{-10pt}

\centering
\begin{tabularx}{\linewidth}{l|c|ccc}
\toprule
\multirow{2}{*}{\textbf{Metrics (\# Trans.)}} & \multirow{2}{*}{\tool} & \multicolumn{1}{c}{$v_1$: w/o} & \multicolumn{1}{c}{$v_2$: w/o} & \multicolumn{1}{c}{$v_3$: w/o} \\
                                &                        & input pairs         &  refinement     & assessment    \\

\midrule
compilable                       & 95     & 87 (-8.42\%)     & 93  (-2.10\%)   & 95  (0.00\%)   \\
\textgreater{}0\% generalizable   & 83     & 73 (-12.04\%)    & 82 (-1.20\%)   & 70  (-15.66\%)   \\
\textgreater{}75\% generalizable & 81     & 66  (-18.51\%)   & 75  (-7.40\%)   & 59  \textbf{(-27.16\%)}   \\
100\% generalizable              & 72     & 58 \textbf{{(-19.44\%)}}    & 61 \textbf{{(-15.27\%)}}     & 56 {{(-22.22\%)}}     \\
\bottomrule
\end{tabularx}
    \vspace{-5pt}
\end{table}

\subsubsection{Result.}
As shown in Table~\ref{tab:rq-ablation}, removing additional input pairs ($v_1$) led to a 19.44\% decrease in generating 100\% generalizable transformations.
This suggests that additional input pairs effectively mitigate the overfitting problem caused by the limited examples in PBE~\cite{DBLP:conf/popl/Gulwani11, DBLP:conf/icse/PanLNGLK21, DBLP:conf/fmcad/AlurBJMRSSSTU13}, helping \tool generate more generalizable transformation.

\editedA{
Similarly, disabling the refinement step ($v_2$) reduced 15.27\% input transformations that generalize to 100\% prepared inputs. 
This indicates that some generated transformations have minor issues and can be refined by excluding irrelevant code.
Besides, disabling the assessment step ($v_3$) decreased 22.22\% input transformation generalizable to 100\% inputs. 
This suggests that, even with additional input pairs and refinement, few 100\% generalizable transformations can be generated, and random selection may miss them. 
The assessment step is necessary to rank the most generalizable function.
}

\begin{answertorq}
All three \todo{designs} contribute to the effectiveness of \tool in generating generalizable transformations. The assessment procedure contributes the most, and additional example input pairs contribute similarly.
\end{answertorq}

\subsection{RQ4: Usefulness of Input Transformations}\label{sec:rq-usefulness}
\subsubsection{Experiment Setup.}
In this RQ, we integrated the generated input transformations into MTCs to construct generalized MRs and measured how well such MRs enhanced test adequacy.
This demonstrated the practical usefulness of \tool's transformations in enhancing test adequacy.

\textbf{New Test Cases Construction.}
We applied generalized MRs to the automatically generated source inputs introduced in Section~\ref{sec:experimental-source-inputs} to obtain a set of new test cases (denoted as~\tranMRBasedTestSuite).
We compare such test cases against two baselines: 
(i) the developer-written test cases (i.e., MTCs) (denoted as \developerWrittenTestSuite) and (ii) 
test cases based on the LLM-generated source and follow-up input pairs (denoted as \llmInputPairTestSuite). Specifically, we combined the prepared source inputs (Section~\ref{sec:experimental-source-inputs}) with valid follow-up inputs generated by \lm{}$^+$ which performed the best in RQ2 (Section~\ref{sec:rq-transforamtionEffectivenss}).
Considering generalized MR based test cases and LLM-generated input pairs based test cases are \todo{extended} from developer-written existing test cases, we followed \citet{mrscout}'s practice to analyze the test adequacy improvement on top of developer-written test cases. 

\textbf{Metrics.} 
We measured test adequacy using two metrics: (i) Line Coverage -- percentage of code lines in target classes executed, and (ii) Mutation Score -- percentage of mutants killed by test cases.

\textbf{Mutation Testing:} 
We employed Pitest~\footnote{https://pitest.org/} to conduct mutation testing. 
Each MR only focused on one or two methods under test in the target class.
To include the covered lines or killed mutants in the methods intransitively\jr{indirectly?}\congying{I am not sure which is better} invoked by MR-involved methods for comparison, we employed Pitest to generate mutants targeting all methods in a target class.
Finally, Pitest successfully generated \todo{4,388} mutants for \todo{45} target classes covered by \todo{88} MRs in the dataset (Section~\ref{sec:dataset}). Pitest failed for the other 12 MRs' classes because of environmental issues (e.g., conflict dependencies).

\subsubsection{Result.}
\begin{table}[h!]
\footnotesize
    \caption{Enhancement of test adequacy from generalized MR based test cases (\tranMRBasedTestSuite) on top of developer-written (\developerWrittenTestSuite) and LLM-generated input pairs (\llmInputPairTestSuite) based test cases}\label{tab:rq-usefulness}
    \centering
    \vspace{-10pt}

\setlength{\tabcolsep}{4pt}
\begin{tabularx}{\linewidth}{l|ccc|ccc}
\toprule
\multirow{3}{*}{\textbf{Metrics}} &
\multicolumn{3}{c}{\textbf{VS. \developerWrittenTestSuite}} & 
\multicolumn{3}{c}{\textbf{VS. \developerWrittenTestSuite+\llmInputPairTestSuite}} \\
\cmidrule(l{1pt}r{1pt}){2-4} 
\cmidrule(l{1pt}r{1pt}){5-7}  
& \textbf{\developerWrittenTestSuite} & \textbf{\developerWrittenTestSuite+\tranMRBasedTestSuite} & \textbf{Improve.}
& {\textbf{\developerWrittenTestSuite+\llmInputPairTestSuite}} & {\textbf{\developerWrittenTestSuite+\llmInputPairTestSuite+\tranMRBasedTestSuite}} & \textbf{Improve.}\\ 
\midrule

Line Coverage  & 0.2373 & 0.2625 & +10.62\% & 0.2588 & 0.2698 & +4.25\% \\
Mutation Score & 0.1322 & 0.1572 & +18.91\% & 0.1710 & 0.1807 & +5.67\% \\
\bottomrule
\end{tabularx}

\end{table}

As shown in Table~\ref{tab:rq-usefulness}, compared to developer-written MTCs (\developerWrittenTestSuite), incorporating new test cases constructed from generalized MR (\developerWrittenTestSuite+\tranMRBasedTestSuite{}) increased the line coverage by 10.62\% and the mutation score by 18.91\%. 
This suggested that \tool could enhance the test adequacy by integrating high-quality test oracles (i.e., output relation of the encoded MR) with a diverse set of potential test input pairs of the MR (\tranMRBasedTestSuite{}).
Although the developer-written test inputs hard-coded in MTCs were carefully crafted and invaluable, each typically included one pair of test inputs and could not sufficiently exercise the SUT's behaviors.
The new source inputs generated by test generation techniques and the corresponding follow-up inputs enabled by \tool may reach program states not covered by the hard-coded inputs.

Besides, by analyzing the benefit of using \tool (\developerWrittenTestSuite+\llmInputPairTestSuite+\tranMRBasedTestSuite) over the test suite enhanced by LLM-generated valid input pairs (\developerWrittenTestSuite+\llmInputPairTestSuite), we could still observe 4.25\% and 5.67\% improvements in the line coverage and the mutation score, respectively. 
This suggested that even if an LLM could act as a black-box transformation to generate some valid source-followup inputs and reach more execution states of SUT, \tool could generate input transformations that apply to more source inputs and better enhance the test adequacy.

\begin{answertorq}
    Test cases constructed from generalized MRs could achieve 10.62\% and 18.91\% increases in the line coverage and mutation score, respectively, demonstrating generalized MRs' practical usefulness in enhancing test adequacy.
\end{answertorq}

\section{Discussion}
\subsection{Threads to Validity} 

We identified potential threats to the validity of our experiments and have taken measures to mitigate them.

\textbf{Representativeness of Experimental Subjects.}
A potential threat is whether our evaluation findings can generalize to different projects. 
To mitigate this threat, we adopted the criteria from existing studies~\cite{mrscout, DBLP:conf/icsm/Wang0HSX0WL20,DBLP:journals/ese/HuangCXWSPWL22} to select high-quality and well-maintained Java projects as representative subjects (Section~\ref{sec:dataset}) and evaluated our method on these projects. 
Besides, evaluating LLMs with subjects seen during model training (known as the data leakage issue) will make the findings biased~\cite{tanlindataleakage}. 
To mitigate this threat, we collected \todo{MR-encoded test cases} created after the training cut-off date of the experimental LLMs, as described in Section~\ref{sec:dataset}. 

\textbf{Representativeness of Experimental LLMs.}
Our method depends on LLMs, and we also use LLMs as baselines.
A potential threat is whether our evaluation findings based on the selected LLMs are representative. 
To mitigate this threat, we evaluated our method with LLMs from three well-known LLM families, i.e., \gpt{} from OpenAI, \lm{} from Meta, and \dps{} from DeepSeek.
They represent the state-of-the-art code LLMs (according to the {EvalPlus leaderborad}\jr{cite}) that can be deployed with the hardware capacity of our machine, as introduced in Section~\ref{sec:eval-environment-LLMs}.

\textbf{Quality of the Experimental Source Inputs.}
As introduced in Section~\ref{sec:experimental-source-inputs}, we used an LLM to prepare new source inputs to assess the generalizability of generated input transformations. 
Low-quality source inputs may threaten the evaluation validity. 
To mitigate this issue, we employed another SOTA code LLM (i.e., \qw{}) which is not the experimental subject to prepare the source inputs. \jr{still strange. can revise when you have time}
We then use the ground truth input transformations to filter out invalid source inputs. 

\textbf{Quality of Ground Truths.}
Besides directly using developer-written input transformations in MTCs (if available) as ground truths, we also manually prepared ground truths for MTCs without input transformations. 
There is a potential threat regarding the quality of our prepared ground truths.
To mitigate this threat, two authors (PhD students) proficient at MT and with more than four years of Java programming experience implemented the ground truths after understanding the intention of the SUTs and the encoded MRs.
Specifically, a ground truth was developed by one participant and reviewed by the other until a consensus was reached.
Furthermore, the developed ground truths are validated against the original source input. 

\editedA{
\subsection{Distinct Advantages of MR-based Tests in Fault Detection}
\textbf{Detecting faults in ``non-testable'' programs.}\jr{Does the term ``non-testable'' exist? If not, avoid using it.}\congying{y, exisitng papers use ``non-testable''. This word is too strong, but I haven't got better one. So, i use the symbol ``''.}
MR-based tests offer distinct advantages in validating \textit{non-testable programs} whose expected outputs for given inputs are hard to specify~\cite{2016-segura-tse,chen2018metamorphic}. The usefulness of MR-based tests in detecting such faults for such programs has been reported in studies~\cite{DBLP:conf/issta/MaS00C23, DBLP:journals/tse/SeguraPTC18, DBLP:conf/kbse/ChenJX21, DBLP:journals/tse/LiuKTC14}. \tool targets encoded MRs for testing Java classes. 
We provide two examples to illustrate this advantage. 
}

\editedA{
As shown in Figure~\ref{lst:PUT-AES}, the class \code{AES} contains methods to encrypt and decrypt a string.
The \code{encrypt} function includes a fault: it mistakenly encrypts the \code{secret} argument instead of the intended \code{source}.
However, the expected string literal after encryption is difficult to specify. This makes it difficult to construct an explicit test oracle based on the expected output to effectively validate the behavior of \code{encrypt}.
}

\editedA{
We collected developer-written tests, EvoSuite-generated tests (following Xu et al.'s practice~\cite{mrscout}), and LLM-generated tests (following Yuan et al.'s practice and generating tests with GPT-4~\cite{yuan2024evaluating}).\jr{Collect for what?}
As shown in Figure~\ref{lst:Test-AES}, we found that the non-MR test (including developer-written and LLM-generated assertions) only checks that the encrypted string is not null or empty. EvoSuite failed to generate any assertions.
However, this test is weak in validating whether the encryption process is correctly implemented. It ensures only the existence of a non-empty output.
}

\editedA{
In contrast, the developer-written MR-based test validates the encrypted string using an MR: $\mathit{x = AES.decrypt(AES.encrypt(x))}$ --- \textit{IF} an input $x$ is encrypted and subsequently decrypted, \textit{THEN} the final result should be $x$. 
The MR-based test successfully detects the fault in encrypting the \code{secret} argument instead of the \code{source}, while the non-MR tests fail to detect the fault.
}
\jr{Cannot understand this fault}
\congying{emmm? which part is confusing?}

\begin{figure}
  \centering
  \includegraphics[width=.48\textwidth]{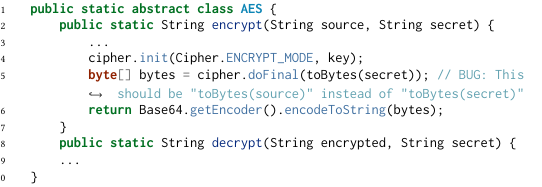}
  \caption{Faulty class \code{AES}{\protect\footnotemark} for encrypting and decrypting}
  \label{lst:PUT-AES}
  \vspace{-5pt}
\end{figure}

\begin{figure}
  \centering
  \includegraphics[width=.48\textwidth]{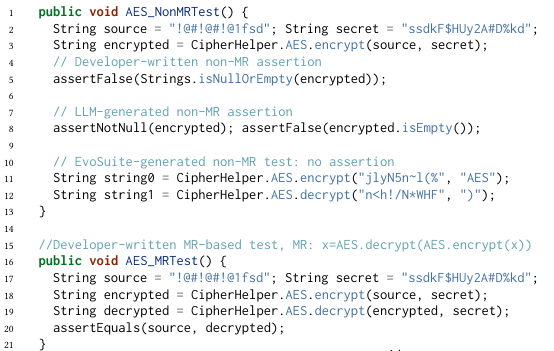}
\caption{Non-MR and MR based tests{\protect\footnotemark} for class~\code{AES}}
\label{lst:Test-AES}
\vspace{-5pt}
\end{figure}

\footnotetext{
Available in the CipherHelperTest.java file from the project \textsc{FlowCI/flow-core-x}
}


\editedA{
Similarly, when testing the \code{getRegistryCenterTime}\footnote{Available in the \code{ZookeeperRegistryCenter} class within the project \textsc{apache/shardingsphere-elasticjob}} function, which is designed to (i) create a new registry entry and return the creation time, or (ii) update an existing entry and return the updating time, it is difficult to determine the expected output because the exact system time depends on when the code is executed.}

\editedA{
The non-MR test, including those with developer-written and LLM-generated assertions, only weakly validates if the registration time is greater than 0 and less than the current system time. EvoSuite failed to generate any assertions. Such a weak non-MR test has a significant limitation:  it fails to validate whether the function correctly handles the update of existing entries. In other words, it does not explicitly check the chronological order of the successive calls.
}

\editedA{
In contrast, a developer-written MR-based test validates the registration time based on an MR: 
\textit{IF} \(t_2=getRegistryCenterTime(key)\) is called after \(t_1=getRegistryCenterTime(key)\), \textit{THEN} \(t_1<t_2\).
When these tests are applied to a faulty \code{getRegistryCenterTime} implementation that returns an unmodified time due to a missing entry update, only the MR-based test can detect this fault, while non-MR tests cannot.
}

\begin{figure}
  \centering
  \includegraphics[width=.48\textwidth]{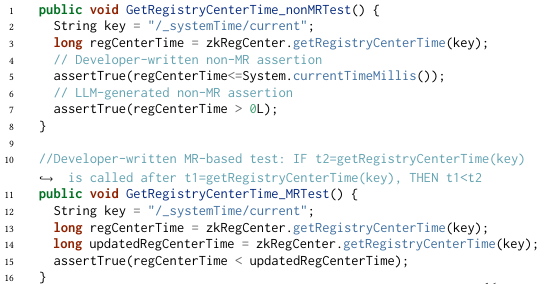}
  \caption{MR-based and non-MR-based tests{\protect\footnotemark} for \code{ZookeeperRegistryCenter}}
  \label{lst:Test-systemTime}
  \vspace{-5pt}
\end{figure}



\footnotetext{
Available in the ZookeeperRegistryCenterQueryWithoutCacheTest.java file from the project \textsc{apache/shardingsphere-elasticjob}
}




\editedA{
\textbf{Exercising a wide range of inputs even with a single MR.}
Another advantage of MT is that \textit{one MR can be applied to a wide range of automatically generated test inputs (known as source inputs) to exercise various program behaviors}. 
}

\editedA{
Our experimental results show that useful tests can be constructed from MRs by leveraging \tool's input transformations and the wide range of generated test inputs. The constructed tests help improve test adequacy with an increase of 10.62\% and 18.91\% in line coverage and mutation score, respectively. 
The results are in line with that reported by Xu et al.'s study~\cite{mrscout}, which observes an increase of 52.10\% and 82.80\% in line coverage and mutation score, respectively, by MR-based tests over EvoSuite-generated tests. 
Even compared with the combination of developer-written and EvoSuite-generated tests, MR-based new tests exclusively covered 113 (+6.95\%) mutants and killed 88 (+7.93\%) mutants. 
This is because, although EvoSuite can generate many inputs, it fails to generate effective test oracles that can identify incorrect outputs and solve the setups to trigger target programs. In contrast, MR-based tests combine high-quality oracles with diverse inputs and leverage developer-written setups for triggering target programs, resulting in higher test adequacy.}
\scc{I cannot follow the meaning of triggering specific programs. Are we talking about program behaviors or program paths or program elements? The last two sentences are difficult to parse. I cannot follow their meaning and the messages.}

\section{Related Work}

\congying{TODO: check PC list, and cite related papers}

\subsection{Automated Identification of MRs.} 

Identification of proper MRs is a key step in applying MT to specific SUTs.
To efficiently identify MRs, many automated approaches have been proposed. 
~
Earlier approaches identify MRs based on a set of predefined patterns \cite{segura2018metamorphic,zhou2018metamorphic}. 
~
\citet{zhang2014search} and \citet{zhang2019automatic} proposed search-based approaches to inferring MRs. 
\citet{DBLP:conf/iccS/TsigkanosRMK23} proposed to use LLMs to identify variable relation and input transformation in scientific software.
These approaches mainly synthesize MRs for specific domains. 
~
\citet{DBLP:journals/corr/abs-2401-17019} proposed an approach to generating executable MRs from requirements specifications using LLMs, but it still requires human effort to implement supportive functions. 
~
Recently, \citet{mrscout} explored a new source to automatically derive MRs.
They synthesize MRs from existing test cases where domain knowledge is embedded.
This served as an effective approach to reusing many encoded MRs. 
~
Such encoded MRs are prevalent, but over 70\% lack an input transformation function to support reusing them on more source inputs. 

{To reuse these invaluable MRs, in this paper,} we propose \tool to generate input transformation functions for such MRs.
Integrated with the input transformations, these MRs are found helpful in enhancing test adequacy in our evaluation.

\subsection{LLMs for Test Generation.}

Researchers explored various LLM usages for test generation. 
~
\citet{arxiv23_FDUGPT4TestGen} studied the performance and limitations of ChatGPT in unit test generation.
~
\citet{ICSE24_Fuzz4All} built a fuzzer using LLMs as a generator of realistic test inputs and an engine for mutation.
~
\citet{TSE24_CompareGPTxSBST} compared the effectiveness of ChatGPT and Evosuite in unit test generation. 
~
\citet{ICSE23CodaMosa} and
~
\citet{arxiv24_covermorebranch} tried to promote the coverage of the tests generated by LLMs.
~

{Different from these works,} \tool does not use LLMs to generate tests directly.
Instead, it generates the input transformation for the encoded MRs and reuses such MRs to enable more tests. 
In fact, using LLMs to generate correct and effective oracles and produce a large number of tests is found challenging~\cite{arxiv23_FDUGPT4TestGen}. 
In comparison, we reuse the human-written oracles in the encoded MRs, which are generally more reliable than LLM-generated oracles.
Besides, MRs can be integrated with test input generation tools to produce {abundant} tests.

\subsection{Enhancing LLMs for Code Generation.}
LLMs are found powerful in code generation \cite{alphacode,ChainOfCode}, attracting numerous efforts to enhance the coding ability further.
Some researchers designed more effective strategies of pre-training \cite{deepseekcoder,starcoder,wizardcoder} and fine-tuning \cite{PanGu-Coder2,stepcoder}. 
Researchers also 
prompted LLMs with compilation messages to guide them to revise the generated code \cite{Lever,selfevolve,arxiv23_FDUGPT4TestGen}
or built a coding agent \cite{ligecodeagent} to enhance LLM's code generation ability.
In light of prompting with analogical reasoning~\cite{ICLR24_Analogical}, our work guides LLMs to generate more examples, identify the intention, and finally generate an input transformation matching the intention.
Also, {different from the approaches that rely purely on LLMs,} we enhance the generated input transformation's quality by performing data-flow analysis to exclude irrelevant code segments from LLMs' responses and ranking the generated transformation functions based on validation with the output relation.

\section{Conclusion}\label{sec:conclusion}


This paper presents \tool, an LLM-based approach to generate input transformations for MRs encoded in test cases that lack explicit input relations.
\tool allows these encoded MRs to be reused with new source inputs, enabling the generation of new tests and achieving higher test adequacy.

Experimental results show that \tool can generate effective input transformations, where \todo{72\%} input transformations are generalizable to all prepared source inputs. When integrated with these transformations and new test inputs, encoded MRs increase line coverage by \todo{10.62\%} and mutation score by \todo{18.91\%}, demonstrating the practical usefulness of \tool’s transformations in enhancing test adequacy.

\section{Data Availability}
We have released the code and data at \url{https://mr-adopt.github.io/}.



\bibliographystyle{ACM-Reference-Format}
\bibliography{src/reference}

\end{document}